\documentclass[aps,pre,preprint,nofootinbib,longbibliography]{revtex4-1}
\usepackage[utf8]{inputenc}
\usepackage{amsmath,amssymb,graphicx,bbm,color}
\usepackage[section]{placeins}

\newcommand{\rbf}{\mathbf{r}}
\newcommand{\ebf}{\mathbf{e}}
\newcommand{\pbf}{\mathbf{p}}
\newcommand{\jbf}{\mathbf{J}}

\newcommand{\nablabf}{\mathbf{\nabla}}

\begin{document}
\title{Dispersal and organization of polarized cells: non-linear
  diffusion and cluster formation without adhesion}
\author{G. Nakamura}
\author{M. Badoual}
\author{E. Fabiani}
\author{C. Deroulers}
\email{derouler@ijclab.in2p3.fr}
\address{Universit\'{e} Paris-Saclay, CNRS/IN2P3, IJCLab, 91405 Orsay, France}
\address{Universit\'{e} de Paris, IJCLab, 91405 Orsay, France}

\begin{abstract}
  Experimental studies of cell motility in culture have shown that
  under adequate conditions these living organisms possess the ability
  to organize themselves into complex structures. Such structures may
  exhibit a synergy that greatly increases their survival rate and
  facilitate growth or spreading to different tissues. These
  properties are even more significant for cancer cells and related
  pathologies. Theoretical studies supported by experimental evidence
  have also shown that adhesion plays a significant role in cellular
  organization. Here we show that the directional persistence observed
  in polarized displacements permits the formation of stable cell
  aggregates in the absence of adhesion, even in low-density regimes.
  We introduce a discrete stochastic model for the dispersal of
  polarized cells with exclusion and derive the hydrodynamic limit. We
  demonstrate that the persistence coupled with the cell-cell
  exclusion hinders the cellular motility around other cells, leading
  to a non-linear diffusion which facilitates their capture into
  larger aggregates.
\end{abstract}
\maketitle

The organization of living organisms into structures with complex
patterns remains an intriguing problem in cell biology
\cite{kochRevModPhys1994}. Even more so if the resulting structures
exhibit enhanced properties that alters the normal functioning of
healthy tissues and their surroundings, or even damage them as a
byproduct of their metabolism. That is often the case for cancer cells
which form abnormal tissues with pathological properties such as
uncontrolled cellular growth or metastastic dissemination
\cite{lobjoisCancerTher2017,sottorivaCancerRes2010}. Metastasis
involves the ability to migrate away from the primary tumor to
secondary sites in healthy tissue, either in an individual or collective
level, and constitute an important aspect of cancer research. This
phenomenon belongs to a broader class described by migration and
displacement of cells \cite{friedlNatRevMolCellBiol2009}. Other
important examples include wound healing
\cite{blanpainNatCellBiol2019} and bacterial swarm
\cite{cotterPNAS2017,jeckelPNAS2019,liPNAS2019}.

Unfortunately, the precise details of collective cell dynamics remains 
unclear at this point due to the rich framework of interactions mediated
by cell-cell junctions, adhesion molecules, or emergent dynamical
phenomena \cite{izardJBiolChem2020}. For instance, it has been
observed experimentally that directional persistence prompts coordinated migration patterns even in the absence of
cell-cell junctions in low-dimensional systems \cite{ladouxNatPhys2020},
striking the stance that explicit interactions are necessary for
collective motions of cells. On the other hand, our understanding of
the molecular and biological processes responsible for the movement of
isolated cells have vastly improved in the past decade
\cite{izardJBiolChem2020,weinerCell2012,ronPhysRevResearch2020}. 
The various sorts of cell movement and their frequencies depends,
naturally, on the morphology of the tissue in question. In the case of
glial cells -- the non-neuronal cells in the nervous system -- their
high mobility can lead to a very aggressive type of cancer shall they
evolve into tumor cells (glioma) \cite{wrenschNatRevNeurol2019}. As
the cancer cells infiltrate into 
the healthy tissue, the spatial boundaries that delimit the tumor 
become more diffuse and ultimately reduce the likelihood of complete
removal of the tumor by surgical interventions. The displacement of
isolated glioma cells occurs through polarization, a mechanism that
reorganizes the inner components of the cell and favors the
displacement along a given direction via action of myosin II
\cite{kimPLOS2017, marshallCurrCellBiol2010}. Polarization confers
glioma cells the ability to 
elongate their bodies along a given direction and cross narrow
intercellular spaces and thus travel long distances inside the brain
parenchyma. The elongation can be interpreted as a morphological
change via actin waves \cite{katsunoTrendsCellBiol2017}, that orients
the traction and retraction of the cell membrane.
The long-lasting nature
of the polarization favors the persistence of cell movements and
produces an average directional migration thanks to correlations
between consecutive displacements \cite{yamadaNatRevMolCellBiol2009}. 

Migration of polarized cells consists of the repetition of three
steps. First, the cell polarizes along a given direction. It then
moves along that direction until it eventually comes to a rest; and
finally it depolarizes restoring its original shape. In the absence of
environmental bias, the polarization direction is chosen randomly
which confers an erratic aspect to the cell migration. More
specifically, the migration of polarized cells describes a velocity
jump stochastic process in the category of the correlated random
walker \cite{codlingJRSoc2008}. Adhesion and exclusion -- the most
prominent cell-cell interactions -- arise as more and more cells are
brought into contact with increasing cell concentrations. Adhesion
encompasses the short range attractive forces that are exchanged
between neighbouring cells via cell-cell junctions and other adhesion
molecules. Exclusion summarizes the inability of any two different 
cells to share the same position at the same time. In
practice, exclusion also reduces the overall cellular mobility as
cells become obstacles for displacements, but it does not create, at least
for cells, a net repulsive force. As a result, cell-cell interactions
tend to further reduce the motility of cells in regions with elevated
cellular concentrations inducing a self-trapping effect. The process
triggers the aggregation of cells and separates the system into two
coexisting phase, namely, a gas phase formed by diluted cells and a
liquid phase comprised of large clusters. By now it is well-known that this
phenomenon describes the motility-induced phase separation (MIPS)
and occurs for self-propelled particles (SPP) even in absence of adhesive forces
\cite{tailleurPhysRevLett2008,tailleurPhysRevLett2012,tailleurNJPhys2018,cottinbizonneNatComm2018}. New
phenomenology in MIPS has been unveiled including the ability to
create topological defects populated with bubbles for 2D active Brownian
particles (ABP) \cite{gonnellaPhysRevLett2020}, breakdown of MIPS with
anisotropic SPP \cite{peruaniNatComm2020}, and emergence of
spontaneous velocity alignment forces \cite{puglisiPhysRevLett2020}.
The connection between MIPS and velocity alignment interactions --
whose foundations stem from the seminal work of Vicsek
\cite{vicsekPhysRevLett1995} -- connects two branches of active matter
physics \cite{marchettiRevModPhys2013,peruaniPhysRevLett2012}
characterized by different types of phase transition. The former
usually invokes first-order phase transitions to create the phase
separation in out-of-equilibrium systems due to particle and/or
temperature imbalance \cite{derridaJStatMech2007}, while the later
employs the rapid transmission of fluctuations across particles in
continuous phase transitions as a way to sustain the information flow
required for self-organization
\cite{bialekPNAS2012,cavagnaPhysRevLett2019}.
{
  Lattice-gas particles with non-convex shapes such as crosses
  suppress rotations and induce persistence. At high levels of
  persistence, regions without particles are observed creating a rich
  nonequilibrium phase diagram \cite{merriganPhysRevResearch2020}.  
  }

Here we formulate a discrete ABP model for polarized cells based on
the typical patterns observed in cellular migration. The growth rate
is assumed to be negligible either by considering short time intervals or
by chemical suppression. We limit cell-cell interactions to exclusion
effects in order to examine the interplay between polarization and
cell density and their effects on cellular organization. Similar
premises were also considered before: in absence of depolarization of
cells before direction changes in \cite{hughesPhysRevE2019}, and in 
the ballistic regime for diluted systems \cite{evansJStatMechTheoryExp2020}.
Migration of cancer cells with adhesion has been addressed before and
compared with experiments on gap junctions
\cite{grammaticosPhysRevE2009} and on formation of deformable
aggregates with proliferation \cite{badoualPLOS2020}.
{
  More recently, a data-driven model has shown that the
  density-dependent propulsion with adhesion promotes the rotation of
  the clusters of cancer cells \cite{copenhagenSciAdv2018}. 
  In \cite{ilinaNatCellBio2020}, the authors investigate the role of
  adherens junctions and constraints of the extracellular matrix on
  cell migration and jamming transitions. More specifically, by
  deregulating the expression of E-cadherin they study the transition
  between collective and individual cell migration and tissue invasion
  in human breast carcinoma.
}
Starting from
our bottom-up approach, Monte Carlo simulations show that cells
self-organize into two phases (gas and liquid) with distinct
properties, with particles being exchanged according to a
power-law. Furthermore, we obtain the mean-field diffusion equation
for the gas phase and demonstrate that the diffusion coefficient
acquires negative values near the gas-liquid interface.
The paper is organized as
follows. Sec.~\ref{sec:model} introduces the model and parameters for
polarized cells. The master equation and relaxation time of global
quantities are described in Sec.~\ref{sec:master}. Self-jamming drives
nucleation and aggregation of cells into large clusters even in
low-density regimes. We describe the phenomenon in
Sec.~\ref{sec:cluster}. The effective diffusion coefficient of the
gas-like phase is derived and compared with simulated data in
Sec.~\ref{sec:diffusion}, with concluding remarks in Sec.~\ref{sec:conclusion}.

\section{Model}
\label{sec:model}

\begin{figure}[htb]
  \includegraphics[width=0.70\columnwidth]{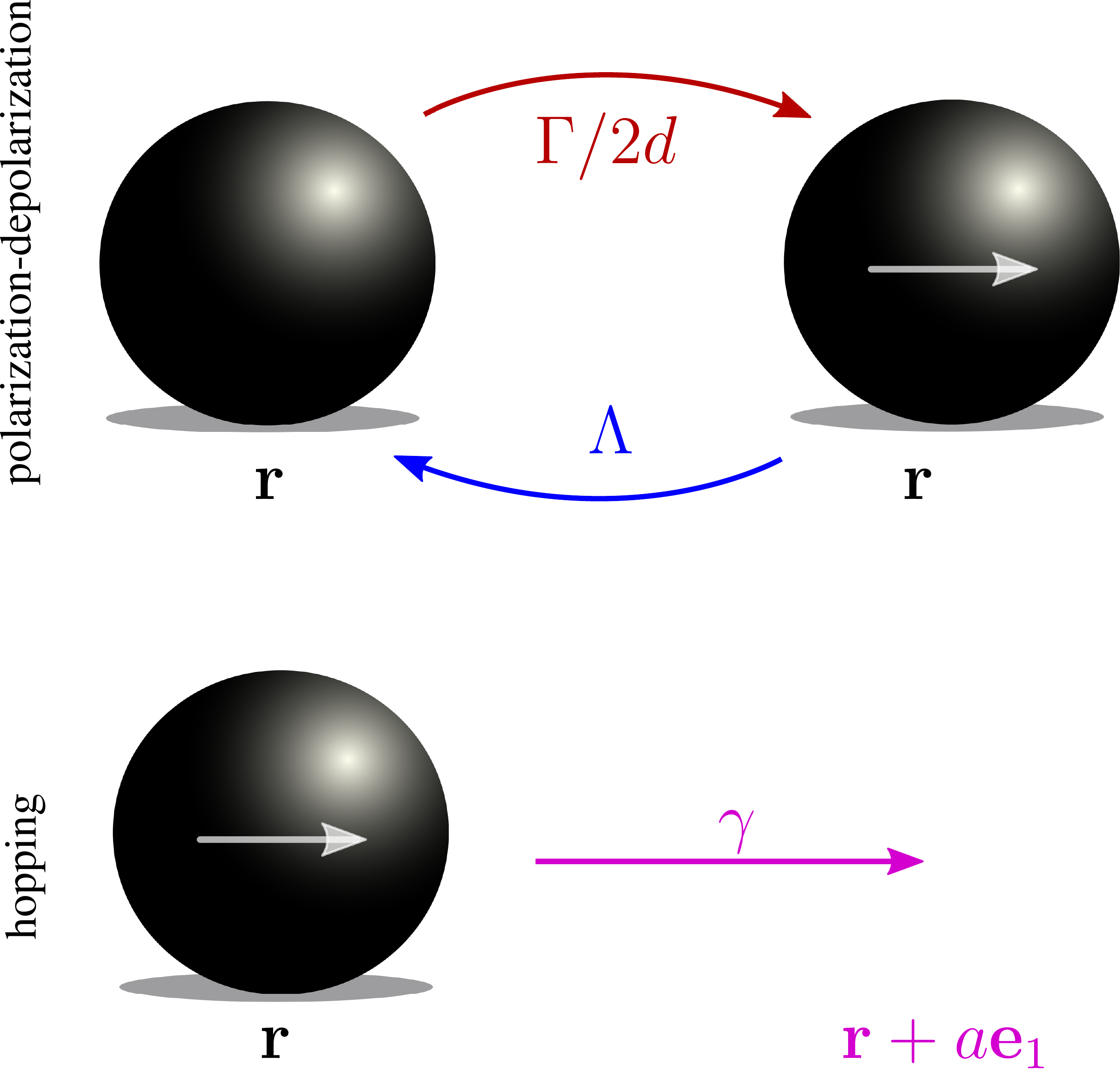}
  \caption{\label{fig:processes} Discrete model. Point-like objects
    represent cells located at the sites of periodic lattice with
    spacing $a$. (top) Cells polarize in a given direction with rate
    $\Gamma/2d$ and return to the non-polarized state with rate
    $\Lambda$. (bottom) Displacement to empty neighbouring sites occur with
    rate $\gamma$ in the same direction of the cellular polarization.} 
\end{figure}

The model describes the displacement of $N$ cells placed in the
sites of a regular square lattice in $d$ dimensions (see
Fig.~\ref{fig:processes}) with volume $V =
L^d$ and orthonormal basis $\{ \ebf_1, \ebf_2, \ldots, \ebf_d\}$.
{
The lattice spacing $a$ is taken to a microscopic length scale of the
order of the typical size of cells. For glioma cells, this scale falls
between 20--40 $\mu$m, although the actual value is immaterial for our
theory.
}
Each site support at most one cell, and each cell can be found in the
non-polarized state or in one of the $2d$ available polarization
states. For the sake of convenience, we associate the non-polarized  
state with the null vector $\ebf_0 = \vec{0}$ while the polarized ones
are either parallel or anti-parallel to coordinate versors, $\pbf_{\pm
  k} = \pm \ebf_k \equiv  \ebf_{\pm k}$. Non-polarized cells lack
discernible mobility and remain in place for $1/\Gamma$ units of time in
average. $\Gamma$ is the polarization rate of the transition from the
non-polarized state to polarized one, and whose direction is sampled
from a uniform distribution. Polarized cells revert to the
non-polarized state with depolarization rate $\Lambda$ or,
equivalently, after a typical time interval $1/\Lambda$. The
correlated displacement requires that cells with polarization $\pbf_k$
can only move along the same direction by $a \ebf_k$ into the next
empty neighbouring site, with movement rate $\gamma$. Furthermore, we
assume the time step $\delta t$ is short enough so that multiple
single cell transitions become unlikely. The initial placement of
cells follows a uniform  distribution with cells polarized in the
direction $\ebf_1$. Table~\ref{tab:params} summarizes the transition
rates.

\begin{table}
  \caption{\label{tab:params} Parameter description. }
  \begin{ruledtabular}
    \begin{tabular}{llc}
      $\gamma$ & displacement rate & \hfill\\
      $\Lambda$ & polarization rate & \hfill\\
      $\Gamma$ & depolarization rate
    \end{tabular}
  \end{ruledtabular}
\end{table}


Given the aforementioned requirements and stochastic rules, one can
implement a simple cellular automaton on a square lattice to produce
numerical simulations. Unless stated otherwise, we also assume
periodic boundary conditions. The cellular automaton goes as
follows. Each time step comprises $N$ consecutive updates of a random
sequence formed by $N$ cells. The random sampling of cells, 
repetitions included, reduces eventual biases caused by improper 
exploration of the phase space (excluding repetitions would reduce the
relaxation time for certain global statistics such as the
concentration of non-polarized cells).

Once a cell is selected, one generates a random number $x \in [0,1)$
from a uniform distribution. 
What happens next depends on the polarization state of the selected
cell. Non-polarized cells polarize for $ x <
(\Gamma\delta t)\mathrm{e}^{-\Gamma \delta t} \approx \Gamma \delta t
$. The new polarization state is obtained by sampling the polarization
direction 
from a uniform distribution. On the other hand, polarized cells may
either move along the polarization axis for $x < \gamma \delta t$
as long as the target site is vacant; or depolarize if $ \gamma \delta
t \leqslant x < (\gamma+\Lambda) \delta t$. It is worth noting that
the time interval $\delta t$ used for single cells appears in all expressions
involving probabilities.
Thus, we can tune the automaton by enforcing that polarized cells
always perform an action within  $\delta t = 1/(\Lambda+\gamma)$ so
that all transition probabilities involve only the ratios between
$\Lambda$, $\gamma$, and $\Gamma$. The choice to tune $\delta t$ based
on transitions of polarized cells comes from observing that, in
general, displacements are more common than morphological changes,
$\gamma > \Gamma,\Lambda$. Alternatively, another common choice takes into
account all transitions rates with equal footing, namely, $\delta t =
1/(\gamma+ \Lambda + \Gamma)$  which tend to be more useful for
general parameter regimes but may requires more steps to reach
equilibrium. Regardless, the choice has no influence on the final 
results as long as one keeps track of $\delta t$. 

Before moving on to the next section, we mention that the assumption
of constant transition rates disregards the natural variability found
in cells. Unlike self-propelled robots and Janus spheres
\cite{cottinbizonneNatComm2018,chatePhysRevLett2010,olivierPhysRevLett2018},
where parameters are strictly controlled, cells age and exhibit
different morphological characteristics. As a result, the distribution
of transition rates depend on specific
characteristics of individual cells. Cellular heterogeneity may
break the assumption of Poisson processes depending on the magnitude
of the variance and number of modes present in the distributions. A remedy to 
this issue contemplates grouping sub-populations of cells sharing the
same physiological characteristics \cite{broederszJRSoc2019}. In our
study, we assume cells belong to the same sub-population and share the
same age.

{\section{The master equation}
\label{sec:master}}

\begin{figure}
  \includegraphics[width=0.9\columnwidth]{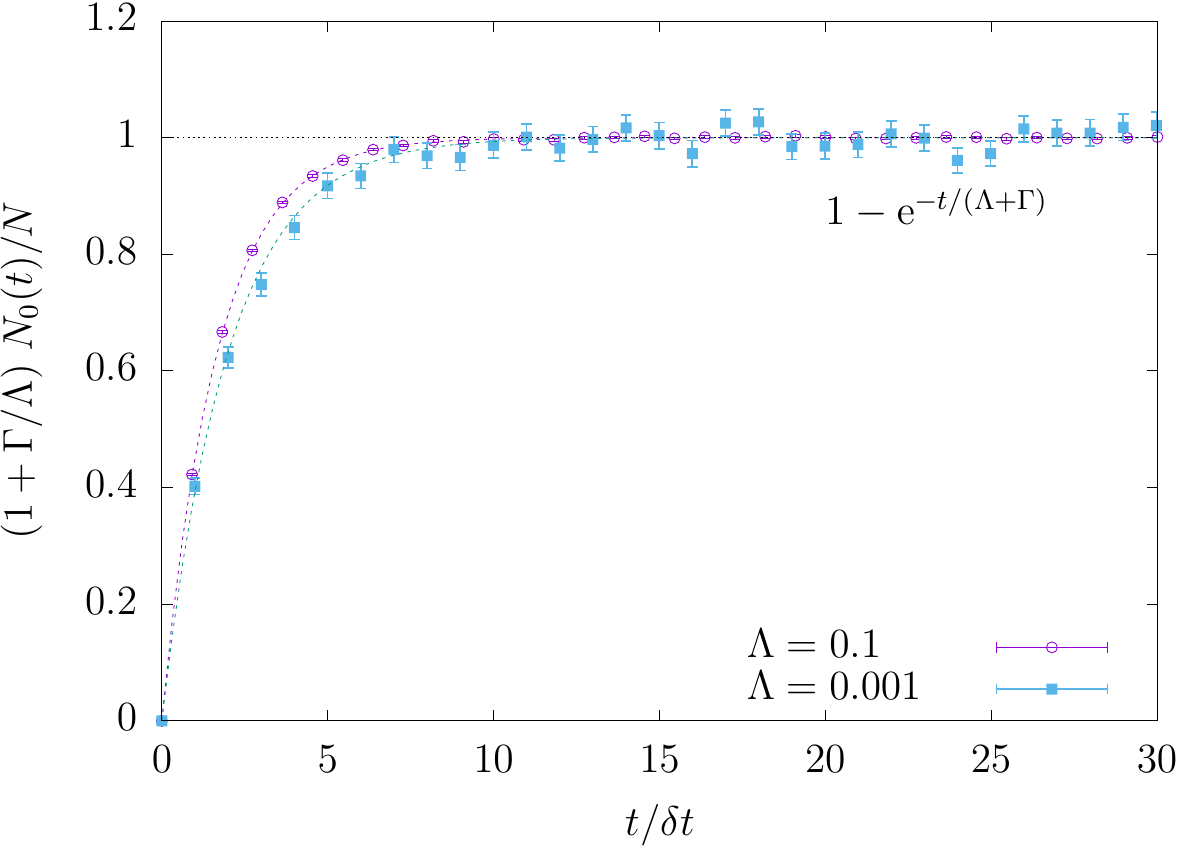}
  \caption{\label{fig:nonpol} Average scaled fraction of
    non-polarized cells along time for distinct values of depolarization
    rate. The parameters are $\Gamma = 1/2$, $\gamma=1$, $L^2=64\times
    64$, $N=100$, with $10^{4}$ Monte Carlo samples. Superimposed
    curves (dashed lines) with
    relaxation time $\tau_0=(\Lambda+\Gamma)^{-1}$. } 
\end{figure}

Let $\rho_k(\rbf,t) \equiv \rho_k(\rbf)$ be the instantaneous density
of cells with polarization $\pbf_k$ at the site $\rbf$, and
$\rho_0(\rbf,t) \equiv \rho_0(\rbf)$ the density of non-polarized
cells. The local cell density $ \rho(\rbf,t) = \sum \limits_{k=-d} ^d
\rho_k(\rbf,t)$ describes the occupation at $(\rbf,t)$ regardless of
the polarization status and it is generally easier to extract from
still images. In addition, the total number of cells $N = \sum_{\rbf}
\rho(\rbf,t) $ is a conserved quantity in the automaton. According 
to the transition rules described in the previous section, and using the 
usual mean-field-like approximation that the presence or absence of 
cells on two distinct lattice sites are statistically independant 
(spatial correlations are neglected~\cite{grammaticosPhysRevE2009}), the 
master equations read
\begin{subequations} 
  \begin{align}
    \label{eq:master1}
    \partial_t\rho_0(\rbf) =& -(\Gamma+\Lambda) \rho_0(\rbf) +  \Lambda \rho(\rbf) , \\
    \label{eq:master2}
    \partial_t \rho_k(\rbf) =& \gamma  \left[1 -  \rho(\rbf) \right]
    \rho_k(\rbf-a\ebf_k) - \gamma \left[1 -\rho(\rbf+a\ebf_k)  \right]
    \rho_k(\rbf) - \Lambda \rho_k(\rbf) + \frac{\Gamma}{2d}
    \rho_0(\rbf),
  \end{align}
\end{subequations}
for $k\neq 0$. Simultaneous occupation of sites is excluded by factors
$1-\rho(\rbf)$. Note that the density of non-polarized cells
(\ref{eq:master1}) lacks contributions from spatial displacements and,
thus, only involves linear terms: it either increases due to the decay
of polarized cells or decreases by transforming into them. In fact,
the total number of non-polarized cells $N_0(t)$ satisfies $(d/dt)N_0
= -(\Gamma+\Lambda)N_0+ \Lambda N$ whose solution reads $N_0(t) =
N/(1+\Gamma/\Lambda) + {c} \textrm{e}^{-t/\tau_0}$ with
characteristic relaxation time $\tau_0 = 1/(\Gamma+\Lambda)$. The
constant $c$ depends on initial conditions whereas the
asymptotic value $N/(1+\Gamma/\Lambda)$ represents the number of
non-polarized cells in the equilibrium. Fig.~{\ref{fig:nonpol}} shows the
evolution of the ratio $N_0/N$ scaled by the factor $(1+\Gamma/\Lambda)^{-1}$
for $\Gamma/\gamma = 1/2$, and $\Lambda/\gamma = 0.1$
or $ 0.001$. In both cases, the initial condition encompasses
$N_1(0)=N$. Cells with lower depolarization rates remain polarized for
longer periods which effectively reduces the asymptotic number of
non-polarized cells. The scaling factor correctly captures the
equilibrium value so that both 
curves converge to unity. However, the curves only overlap around the
equilibrium value suggesting different relaxation times, as
indicated by the superimposed dashed curves
$1-\textrm{e}^{-(\Gamma+\Lambda)t }$, in good agreement
the theoretical predictions for $\tau_0$. 
Of course, the
results for the total number of non-polarized cells should not present
any surprise since in practice it describes a linear problem without
any kind of spatial correlation.

The master equations (\ref{eq:master1}) and (\ref{eq:master2})
describe the stochastic dynamics of polarized cell and are
thus subjected to fluctuations. The equivalent equations for the
mean values $\langle \rho_k(\rbf) \rangle $ are obtained by averaging
the equations over an ensemble of independent replicas. The procedure
leads to contributions $\langle \rho_k(\rbf)\rho(\rbf \pm a\ebf_{k})
\rangle$ which includes two-point spatial correlations between
neighbouring sites. It turns out that the dynamical equations for 
two-points correlations depend on three-points correlations, while the
equivalent equation for three-points correlations requires inputs from
4-points correlations, etc. In short, the displacement of polarized
cells form a set of hierarchical equations for correlations which only
ends at $N$-points. Instead of solving the complete set of equations
one can gain valuable insights on the movement of polarized cells by
studying the problem in the low density regime, where cells are likely
uncorrelated due to the absence of long-range interactions. A coarse
estimate of correlations is given by the mean-field approximation 
$\langle \rho_k(\rbf)\rho(\rbf \pm a\ebf_{k}) \rangle \approx \langle
\rho_k(\rbf)\rangle \langle \rho(\rbf\pm a\ebf_k)\rangle$ which
neglects fluctuations entirely. Note that the mean-field
approximation does not affect correlations involving the cellular
polarization as in absence of additional cell-cell interactions they are
only affected by the depolarization rate.
In what follows, we drop the angular brackets to simplify the
notation.

In addition to the mean-field approximation, it is convenient to
introduce two new sets of variables derived from $\rho_k(\rbf,t)$ that
capture the isotropic nature of the procedure that select new polarization
states. This property ensures that, in practice, the local cell densities
exhibit reflection and rotation symmetries once the distribution of
polarization states reaches its equilibrium regime, characterized by
the short relaxation time $\tau_0$ (see Fig.~\ref{fig:nonpol}). Thus,
the density $\phi_k(\rbf,t) \equiv \phi_k =\rho_k+\rho_{-k}$ and the
corresponding polarization current density $J_k(\rbf,t) \equiv J_k =
\gamma( \rho_k -\rho_{-k})$ are more suitable quantities to describe
the problem:
\begin{subequations}
  \begin{align}
    \label{eq:master11}
    \partial_t \phi_k & = -\Lambda \phi_k + \frac{\Gamma  \rho_0}{d} -a
    \partial_k\left[ (1-\rho) J_k\right]
    + \frac{a^2 \gamma}{2} \left[ (1-\rho)\partial_k^2  \phi_k
    + \phi_k\partial_k^2  \rho \right]
    + \mathcal{O}(a^3),\\
    \label{eq:master12}
    \partial_t J_k & = - \Lambda J_k - a \gamma^2 \partial_k \left[
      (1-\rho) \phi_k \right] +\mathcal{O}(a^2),
  \end{align}
\end{subequations}
with (\ref{eq:master1}) unchanged.
One may further simplify (\ref{eq:master11}) and (\ref{eq:master12})
once enough time has passed to account for the repolarization of
individual cells, $t \gg \tau_0$. The quasistatic approximation of
(\ref{eq:master1}) gives $ \phi_0(\rbf) = (1+\Gamma/\Lambda)^{-1}
\rho(\rbf)$ while the isotropy of space ensures that on average
$\phi_k $ behaves similarly in all directions, $\phi_k(\rbf) \approx
(1/d) \sum_{m=1}^d \phi_m(\rbf) =
d^{-1}(1+\Lambda/\Gamma)^{-1}\rho(\rbf)$.  Combining
(\ref{eq:master11}) for $k=1,2,\ldots, d$ produces
\begin{subequations}
  \begin{align}
    \label{eq:master21}
   \partial_t \rho & = -a \nablabf \cdot [ (1-\rho) \jbf ] +
   \frac{a^2
     \gamma}{2d}\left(\frac{1}{1+\Lambda/\Gamma}\right)\nabla^2\rho
   +\mathcal{O}(a^3) ,\\
   \label{eq:master22}
   \partial_t \jbf & = -\Lambda \, \jbf -\frac{a \gamma^2}{d} \left(
   \frac{1}{1+\Lambda/\Gamma}\right) \nablabf[(1-\rho)\rho] +\mathcal{O}(a^2).
  \end{align}
\end{subequations}

\begin{figure}
  \includegraphics[width=0.9\columnwidth]{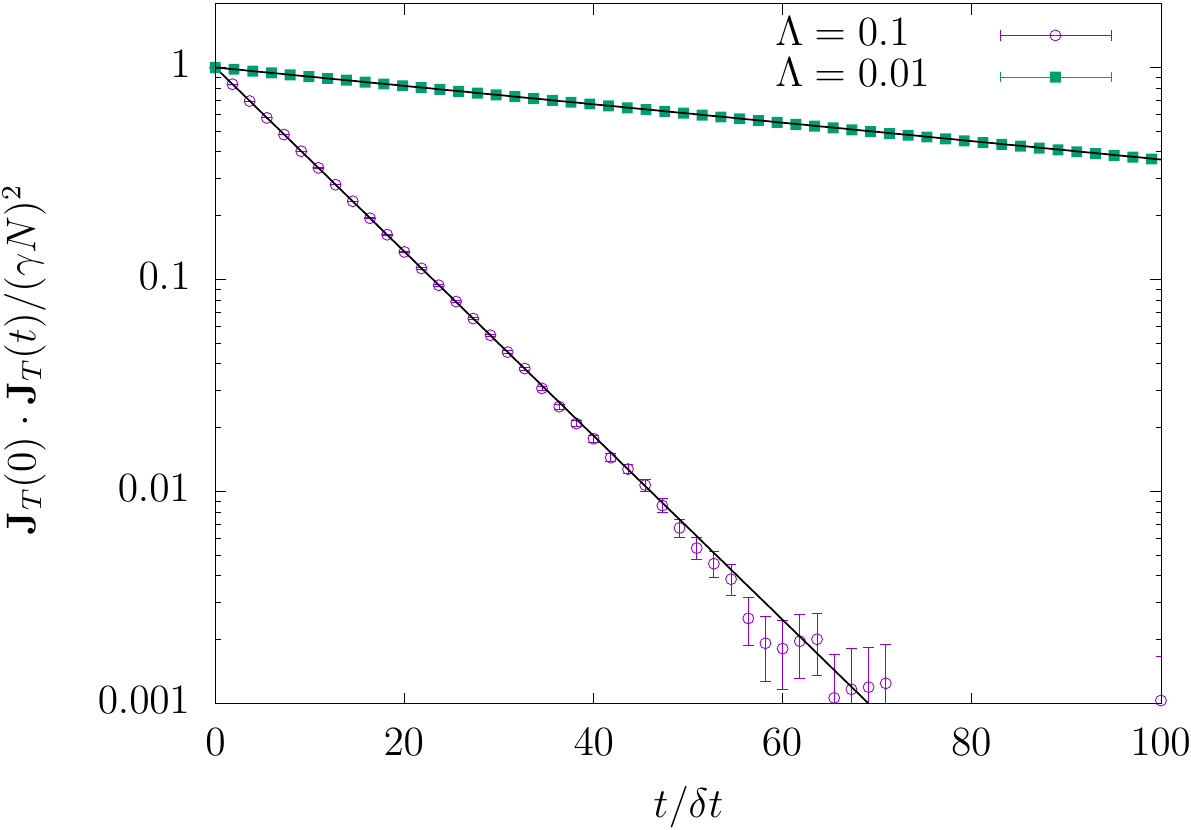}
  \caption{\label{fig:current} Projection of the instantaneous total
    current $J_T(t)$ on $J_T(0)$. The parameters are $N=100$,
    $L=64$, $\Gamma = \gamma/2$ for $\Lambda/\gamma =0.1$ (circles) or
    $0.01$ (filled squares) in logscale, with $10^4$ Monte Carlo
    samples. The solid lines are given by $\exp(-\Lambda t)$. $50\%$
    of the data is omitted for clarity.}
\end{figure}

The equation for total current $\jbf_T(t)$ can be obtained by integrating
(\ref{eq:master22}) over the volume $V$ and using 
the divergence theorem:
\begin{equation}
  \frac{d \jbf_T}{dt} = -\Lambda\, \jbf_T -\frac{a \gamma^2}{d} \left(
   \frac{1}{1+\Lambda/\Gamma}\right) \oint{\rho(1-\rho) d\mathbf{S}_d}
   + \mathcal{O}(a^2).
\end{equation}
For periodic boundary conditions, the surface integral vanishes and
$|\jbf_T | = | \jbf_T(0)|  \textrm{e}^{-\Lambda t}$ describes a simple decay
process with relaxation time $\tau_1 =
\Lambda^{-1}$. Fig.~\ref{fig:current} depicts the evolution of the
total current for distinct depolarization rates $\Lambda/\gamma=0.1$
and $0.01$, with the remaining parameters the same as in 
Fig.~\ref{fig:nonpol} and  initial condition $\jbf_T(0)=\gamma N
\ebf_1$. The solid lines are given by $\textrm{e}^{-\Lambda t}$ and
show good agreement with the simulated data. Deviations between the
curves and data become more significant as $| \jbf_T| / \gamma N$
approaches the typical error  $1/\sqrt{M}$ associated with $M$
independent Monte Carlo samples.

\section{Cluster formation}
\label{sec:cluster}

\begin{figure}
  \includegraphics[width=0.9\columnwidth]{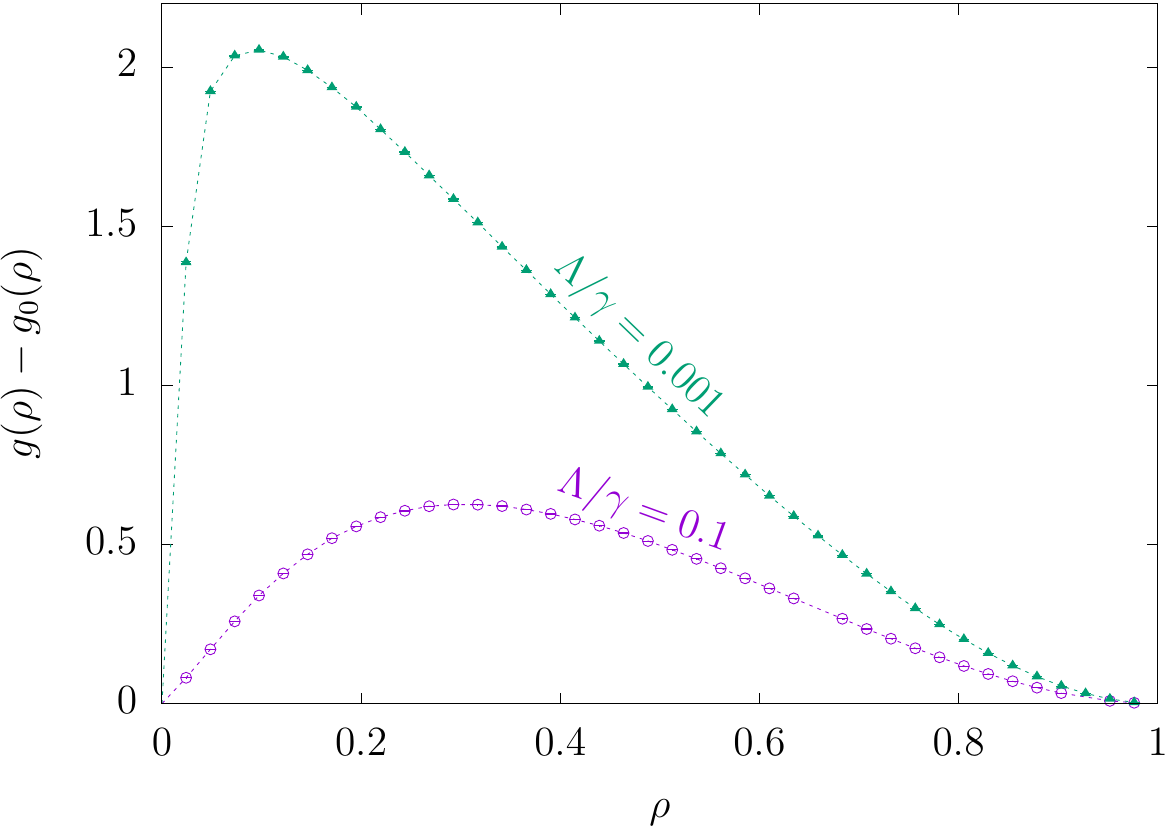}
  \caption{\label{fig:corr} Deviation in the number of adjacent cells
    in the polarized cell model compared to the random walker in $2d$
    ($V=64^2, \Gamma/\gamma = 1/2$) for several levels of cell
    density. }  
\end{figure}

Although there are spatial contributions in (\ref{eq:master21}) and
(\ref{eq:master22}), the global quantities $N_0$ and $J_T$ are
described solely in terms of the polarization-depolarization
transitions. Translation symmetries often mask contributions coming
from the various lattice sites so the local equations for cellular
density require a more careful examination. In fact, local spatial
correlations appear not only with increasing $N$ but also with
decreasing values of $\Lambda$ as the persistence of cellular
migration eventually creates long-lasting collisions between different 
cells. Fig.~\ref{fig:corr} shows the deviation of the average number
of adjacent cells $g(\rho)$ in the polarization model in comparison
with the uncorrelated random walker $g_0(\rho)=2d \rho$ in a periodic
square lattice, both in equilibrium. The average numbers of
adjacent cells are obtained from the pair distribution which in turn
maps the complete spatial correlation between pairs. Thus,
Fig.~\ref{fig:corr} shows that the system undergoes a dramatic change for
vanishing $\Lambda$: cells tend to remain close to each other with
enhanced spatial correlation and reduced mobility, violating the
underlying hypothesis of the mean-field approximation.


\begin{widetext}
\begin{figure}
  \includegraphics[width=0.95\textwidth]{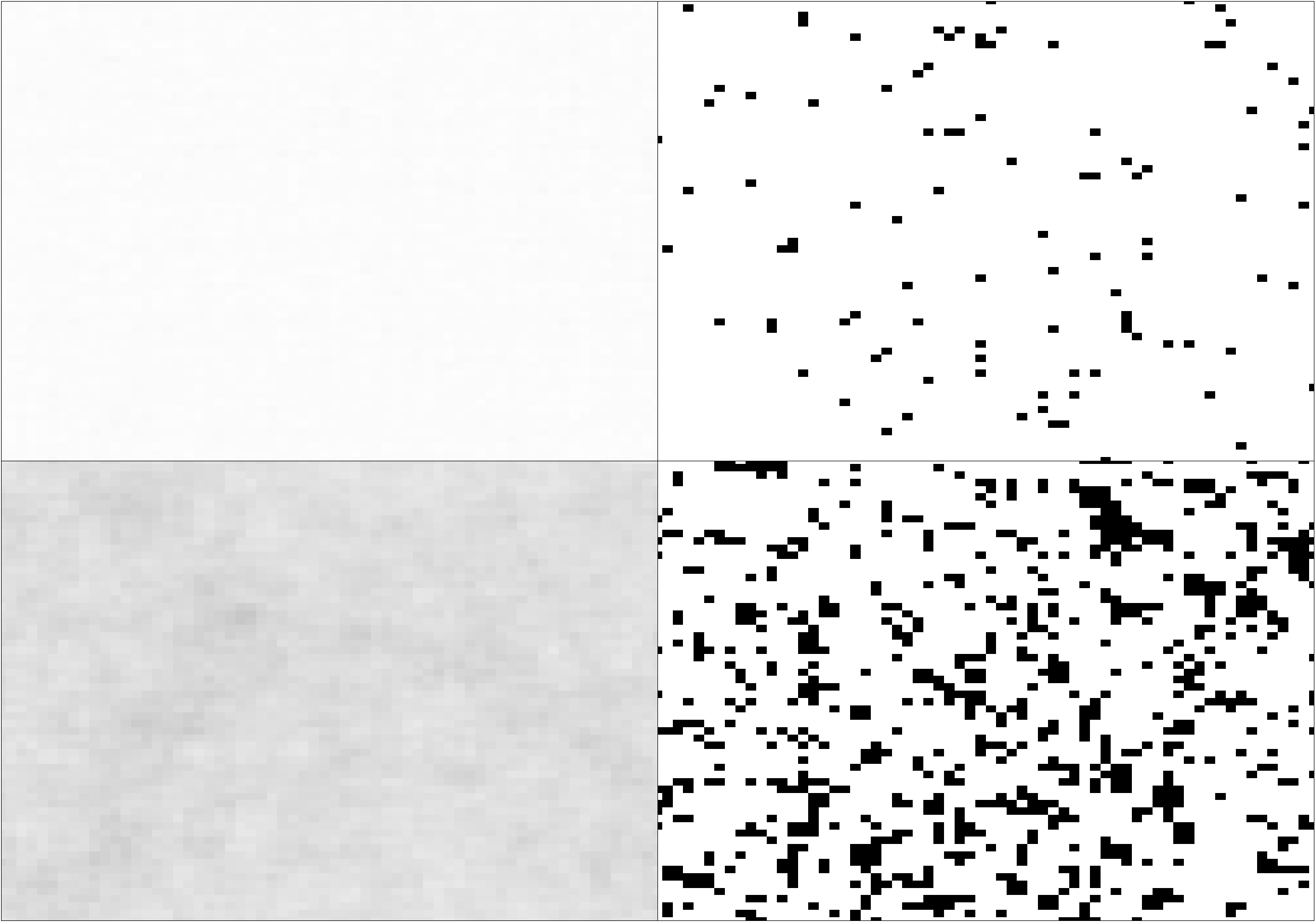}
  \caption{\label{fig:clusters1} Cellular aggregation for large
    depolarization rate. Time average of cell distributions over
    $10^4$ samples (left
    column) with corresponding configuration snapshots (right
    column). $\Lambda/\gamma=0.1$ with $N=100$ cells (top) and $N=800$
    (bottom).  Parameters $V=64^2, \Gamma/\gamma=1/2$ and samples collected after $5\times 10^4$ steps.}
\end{figure}
\end{widetext}

\begin{widetext}
\begin{figure}
  \includegraphics[width=0.95\textwidth]{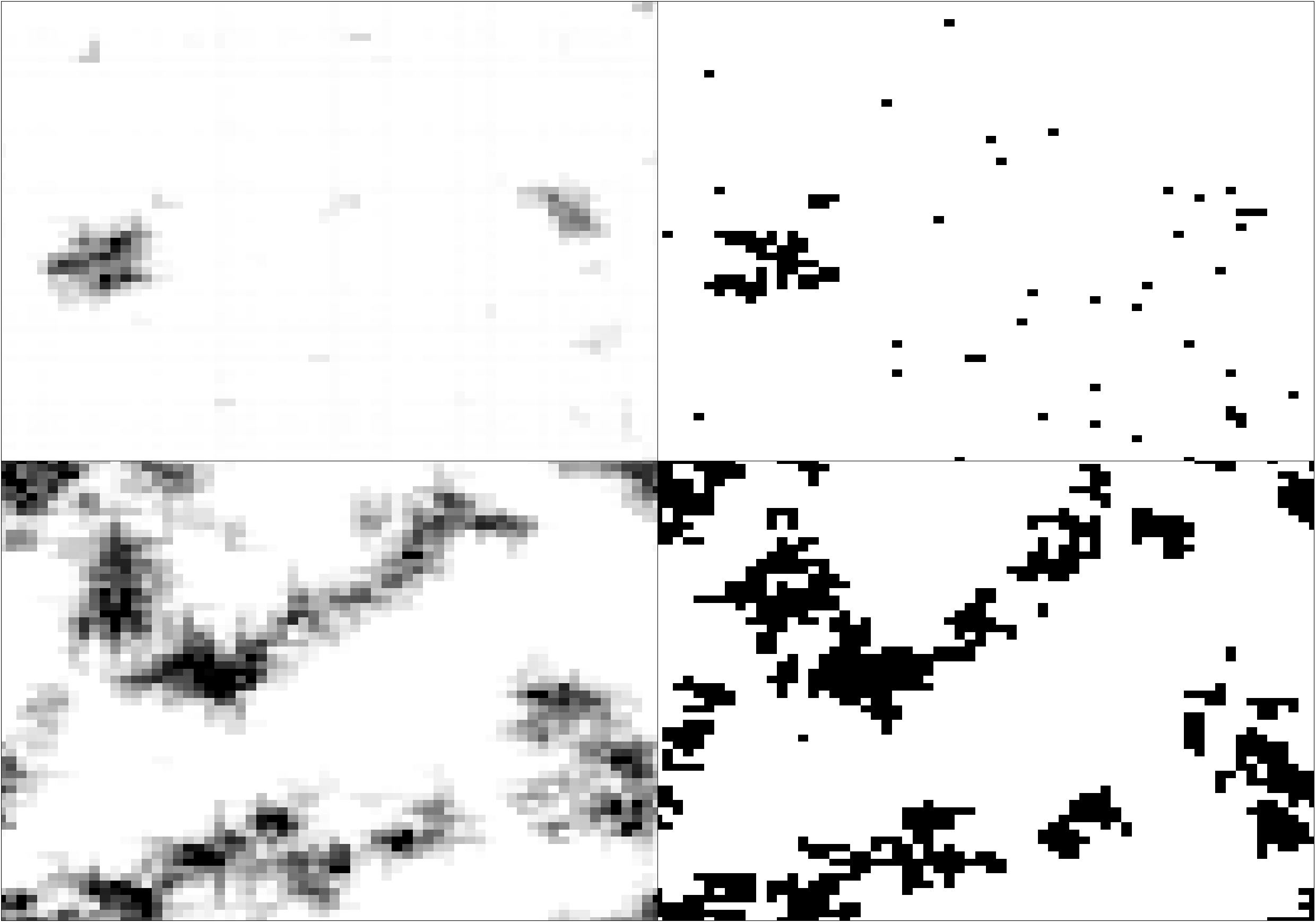}
  \caption{\label{fig:clusters2} Cellular aggregation for low
    depolarization rate. Time average of cell distributions over
    $10^4$ samples (left column) with corresponding configuration snapshots (right
    column). $\Lambda/\gamma=0.001$ with $N=100$ cells (top) and $N=800$
    (bottom). Parameters $V=64^2, \Gamma/\gamma=1/2$. Samples collected after $5\times 10^4$ steps.}
\end{figure}
\end{widetext}

Despite the lack of cohesive forces in the polarization model, cells
may form long-lasting aggregates. Fig.~\ref{fig:clusters1} and
Fig.~\ref{fig:clusters2} illustrate four instances of the
polarization model in $d=2$ with different values of
$\Lambda/\gamma=0.1$ and $0.001$, respectively, for different
concentrations of cells. After a long transient ($5\times 10^4$
steps), the spatial distribution of cells is recorded for each
instance and averaged over time {( left column,
  Fig.~\ref{fig:clusters1} and Fig.~\ref{fig:clusters2})}.  For $\Lambda/\gamma = 0.1$ the
spatial distribution closely resembles the uniform distribution. Small
clusters eventually form but they quickly break apart owing to their
high depolarization-polarization rate. This mechanism ensures the
complete visitation of the lattice and preserves the complete
translation symmetry of the system. However, low depolarization
rates $\Lambda/\gamma = 0.001$ promotes the formation of stable
clusters. 
Cells along the cluster boundaries are still allowed to escape and
dictate the equilibrium condition, namely, the regime in which the
average number of departing and arriving cells are
equal or, equivalently, a vanishing exchange current between the
aggregated and isolated cells.


It is entirely reasonable to link cellular migration with cluster
dynamics and related statistics (e.g. surface area, perimeter, and
centroid). In exchange, one must obtain new master equations for the
target cluster statistics that might or might not be derived from
(\ref{eq:master21}) and (\ref{eq:master22}). The problem with this
approach stems from 
the fact that we only know the basic stochastic rules that govern the
movement of cells, not their collective behavior. We avoid the search
for new equations and instead introduce two familiar quantities to
study the system, namely,
\begin{subequations}
  \begin{align}
    R^2(t_k) & \equiv \left| \sum_{\ell=0}^k \sum_{m=1}^N 
\left[    \rbf_m(t_{\ell+1}) - \rbf_m(t_{\ell})\right] \right|^2 ,\\
    S^2(t_k) & \equiv \frac{1}{N} \sum_{m=1}^N 
    \left| \rbf_m(t_{k+1}) - \rbf_m(t_{k})\right|^2.
  \end{align}
\end{subequations}
$R^2(t)$ is simply the squared displacement or,
alternatively, the square integrated velocity $R^2(t_k) = \delta
t^2 \sum_{\ell,\ell'}^k\sum_{m,m'}^N \mathbf{v}_m(t_\ell) \cdot
\mathbf{v}_{m'}(t_{\ell'}) $. The average $\langle R^2\rangle$ thus
measures the collective dispersion of cells and involves the
autocorrelation between cell velocities. For particles with negligible
autocorrelation times one observes a regular diffusive behavior
$\langle R^2(t)\rangle \approx (2d N) D\, t$ expressing the
relationship between $\langle R^2\rangle$ and the diffusion
coefficient $D$ \cite{thorstenChaos2005}.

\begin{figure}
  \includegraphics[width=0.9\columnwidth]{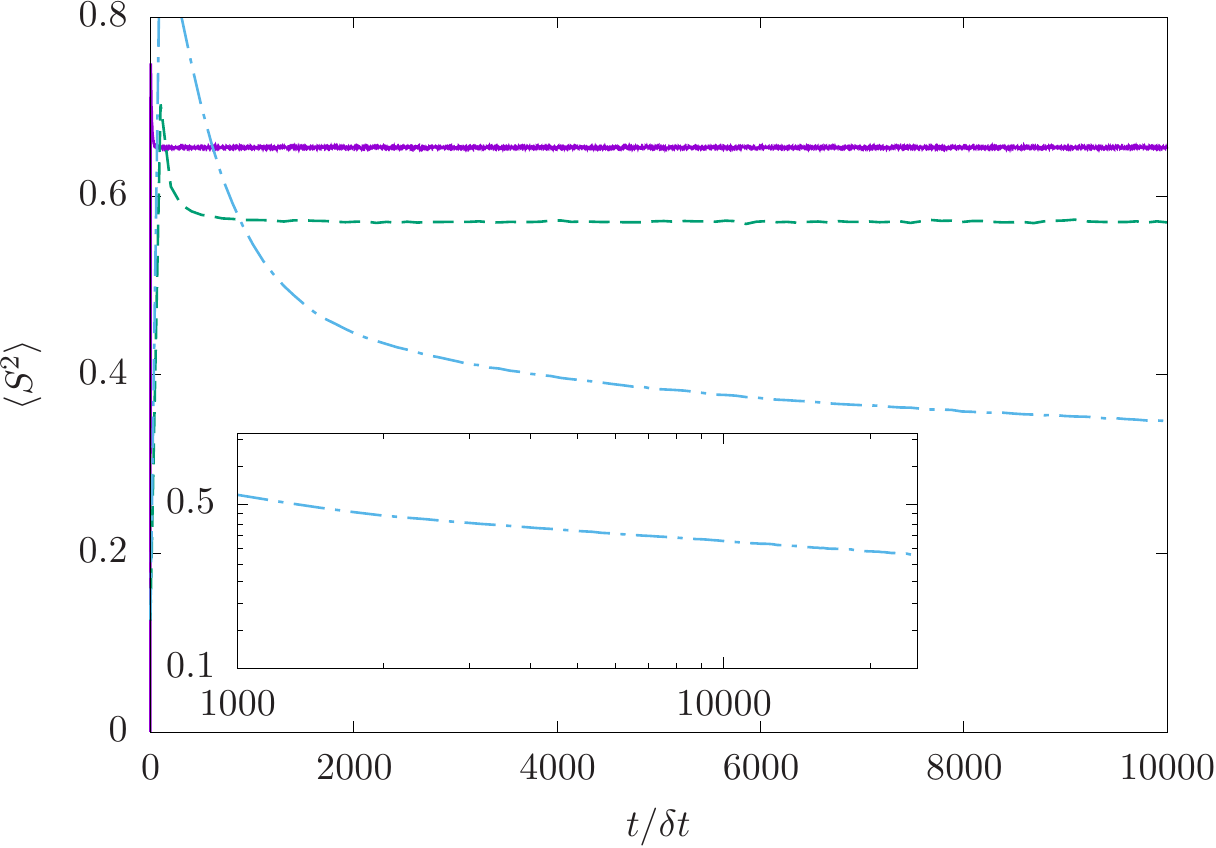}
  \caption{\label{fig:kinetics} Average  one-step square displacement
    per particle. $V=64^2, \Gamma/\gamma=1/2, N=100$ and $10^4$ Monte
    Carlo samples. {Average one-step square displacement} $\langle
    S^2\rangle$ { equilibrates within times of the order of a few thousand steps for cluster-free configurations
    $\Lambda/\gamma=0.1$ and $0.01$ ( --- and --$\,$-- respectively)
    but decays much more slowly, as a an apparent power-law} for
  $\Lambda/\gamma=0.001$ (-- $\cdot$ --) . (inset) log-log scale.
}
\end{figure}

\begin{figure}[htb]
  \includegraphics[width=0.9\columnwidth]{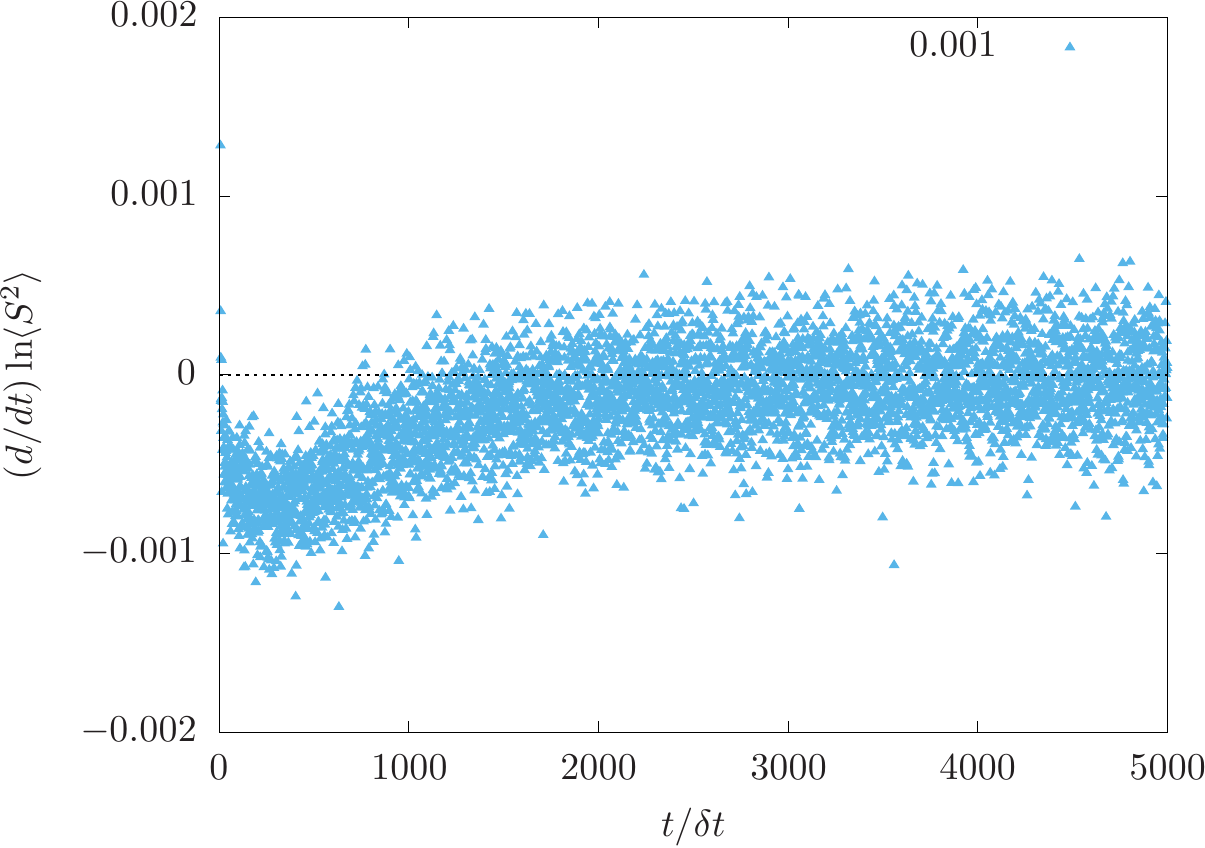}
  \caption{\label{fig:kinetics_relative} Relative variation of one-step
     square displacements. $V=64^2, \Gamma/\gamma=1/2,  N=100$
    and $10^4$ Monte Carlo samples with $\Lambda/\gamma=0.001$. }
\end{figure}

In all cases, cells experiment an explosive initial dispersion --
owning to movement persistence
-- followed by a rapid decrease of the dispersion
rate throughout the lattice. The phenomenon begins shortly after a
time interval $\tau_1$ has passed due to depolarization and it is
enhanced by plastic collisions between cells. A portion of these
collisions disrupt the cellular flow across the lattice producing a
surge in the rigidity of the system (self-jamming effect). In
practice, the presence of jams shift the tissue to a more liquid-like
behavior. The transition from gas to liquid is one example of jamming
transitions, which are believed to play an important role in tumor
development and metastasis \cite{kasJPhysDApplPhys2017} and
epithelial-to-mesenchymal transition \cite{marchettiPhysRevX2016}.
For vanishing ratios $\Lambda/\gamma$,
the longer persistence of cellular displacement enhances the duration
of jams, which in practice become nucleation centers for
clusters. These special structures act as barriers
that entrap arriving particles and may include up to $\mathcal{O}(N)$ cells. 
Once clusters are in place, the slope of $\langle R^2 (t)\rangle$ is
governed by the displacement of their peripheral cells as well as the
relative few unbounded ones. The former encodes the fluctuations of
departing and arriving particles into clusters as indicated by the
light-gray regions at cluster boundaries in
Fig.~\ref{fig:clusters2}.

The behavior of fluctuations generally differs from the bulk
counterpart and it is often accompanied by the emergence of power-laws
\cite{chialvoPhysRevLett2017}. Here, this behavior is captured by the
average one-step square displacement $\langle S^2(t) \rangle$ as
shown in Fig.~\ref{fig:kinetics}. More specifically, $\langle S^2(t)
\rangle$ decays as a power-law for vanishing $\Lambda/\gamma$ in sharp
contrast with the constant behavior observed in cluster-free
configurations. 
{In practice, plastic collisions reduce the overall displacement rate
  while cells escaping from clusters increase it, creating an 
  out-of-equilibrium steady state. }
We can obtain a better assessment of
equilibrium, or quasi-equilibrium for a lack of a better word, by
using the relative variation $(d/dt)\ln \langle S^2(t) \rangle$, which
fluctuates around zero once the bulk part of clusters is in place as
depicted in Fig.~\ref{fig:kinetics_relative}. { Here the derivative is
computed as a forward derivative over one discrete time step. }

\begin{figure}
  \includegraphics[width=0.9\columnwidth]{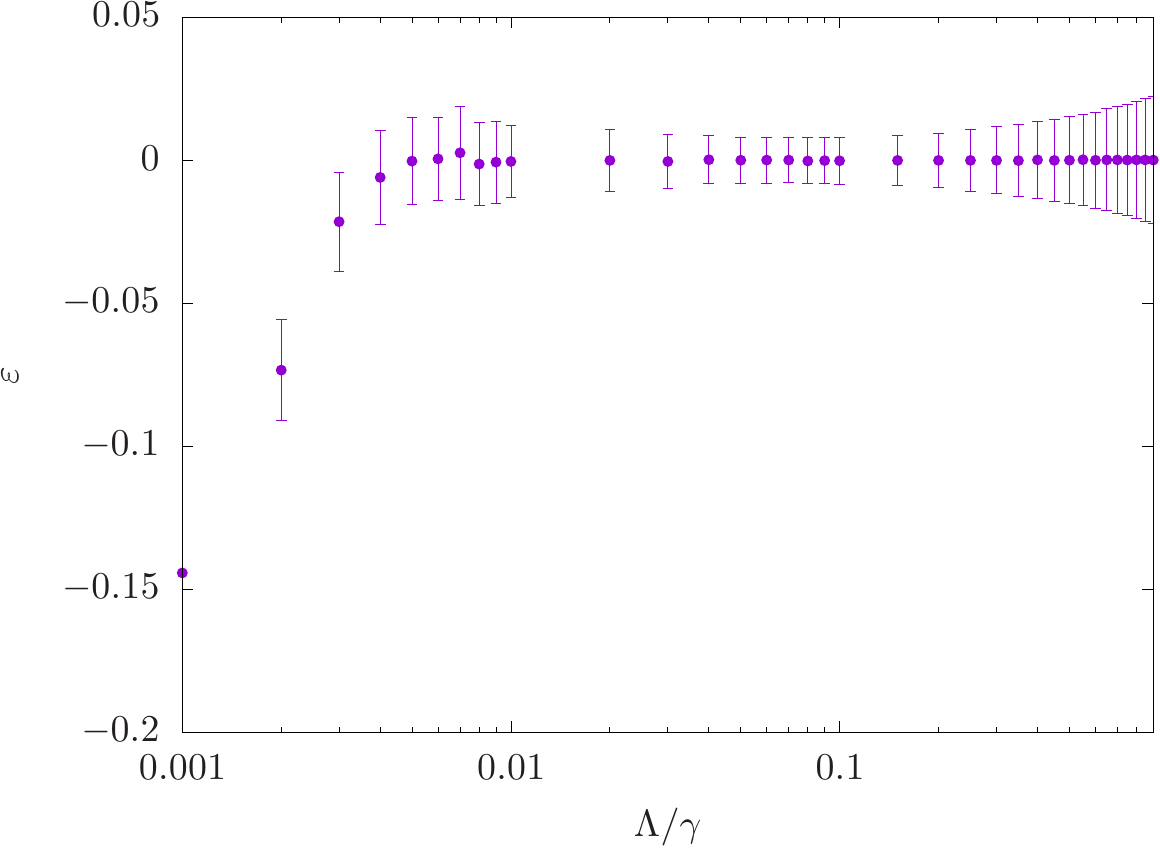}
  \caption{\label{fig:kinetics_depol} Order parameter. The escape rate
    $\varepsilon$ vanishes in the cluster-free regime. $N/V =
    100/64^2$, $\Gamma/\gamma=0.5$.} 
\end{figure}

We can exploit the effective power-law to define the escape rate
$\varepsilon = \partial \ln \langle S^2 \rangle / \partial \ln t $
which vanishes in the free particle regime or acquires a non-null
value in the cluster phase. The values that $\varepsilon$ can
assume in the cluster phase depend on the various transition rates
$(\gamma,\Gamma$ and $\Lambda$) and concentration $N/V$.
Fig.~\ref{fig:kinetics_depol} depicts an example for various values of
$\Lambda$. As expected, $\varepsilon=0$ in the free-particle phase but
acquires negative values for some  $\Lambda <  \Lambda'$  indicating the
emergence of clusters. In Fig.~\ref{fig:kinetics_depol},
we find that $\partial \varepsilon /\partial \Lambda $ diverges for
vanishing $\Lambda$ if the relation $\varepsilon \sim \ln
\Lambda/\gamma$ holds. 

\section{Non-linear diffusion in the diluted regime}
\label{sec:diffusion}


The self-organization of polarized cells into long-lasting clusters
shifts the description of the problem of microscopic entities (cells)
to macroscopic ones (clusters). This occurs because the collective
movement of cells in a given cluster can be perceived as the movement
of the cluster itself. However, the macroscopic equations are not
immediately obvious or known a priori unlike their microscopic
counterparts. Phenomenological equations are good candidates to tackle
the macroscopic approach but might require additional phenomenological
parameters in order to address the various sorts of interactions that
might arise. 

We circumvent the issue altogether by studying the regions located far
away from cluster boundaries. Because clusters capture surrounding
particles, the remaining regions become deprived of cells and
can be treated as a diluted phase. We refer to Fig.~\ref{fig:clusters2}
where the emergence of clusters causes a dramatic reduction of
migrating cells when compared to configurations with higher
depolarization rates. Clusters describe a liquid-like phase
whereas the surrounding regions populated by $N'$ cells represent a
diluted gas phase. As a result, only weak correlations survives between $N' < N$
cells in the diluted gas phase which in turn satisfies the
requirements imposed by the mean-field equations (\ref{eq:master21}) and
(\ref{eq:master22}).

\begin{figure}
    \includegraphics[width=0.9\columnwidth]{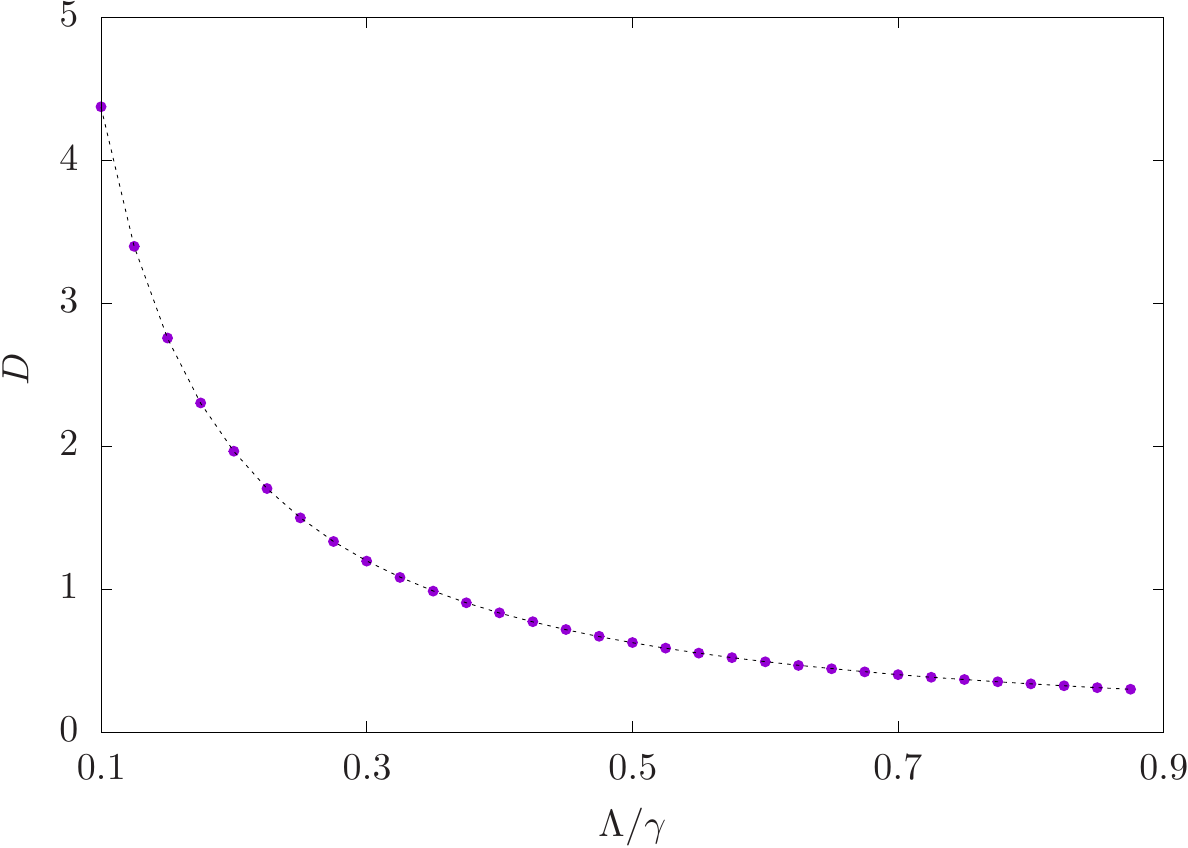}
  \caption{\label{fig:single} Global diffusion coefficient per site
    for a single cell in the periodic square lattice. A least-square fitting
    procedure from the asymptotic behavior of $\langle R^2 \rangle
    \sim 4 D  t$ extracts the global diffusion coefficient $D$
    (filled circles). The dashed line represents the spatial average
    of $D_{\textrm{eff}}$. Parameters $\Gamma/\gamma=0.5, V=64^2$ and
    $10^6$ Monte Carlo samples.  
  } 
\end{figure}

With the above considerations in mind, we can obtain estimates for the
local diffusion coefficient in the diluted gas phase. For that
purpose, it is convenient to rewrite (\ref{eq:master21}) provided the
quasistatic approximation $\jbf \approx - (a  \gamma^2/d \Lambda)
(1+\Lambda/\Gamma)^{-1} (1-2\rho)\nablabf\rho $, obtained from
(\ref{eq:master22}) with $t\gg \tau_1$. Plugging the quasistatic
approximation for $\jbf$  into (\ref{eq:master21}) produces 
\begin{equation}
  \label{eq:master_eff}
  \partial_t \rho =   \frac{\gamma a^2}{2 d}
  \left( \frac{\Gamma}{\Gamma+\Lambda}\right) \nablabf \cdot
  \left[ 1 +\frac{2\gamma}{\Lambda} (1-\rho)(1-2\rho) \right]
  \nablabf \rho .
\end{equation}
Starting from the continuity equation, one quickly identifies
the cell current $\mathbf{j}_{\textrm{eff}} = -D_{\textrm{eff}}
\nablabf \rho$, in close analogy with Fick law, with the effective
diffusion coefficient 
\begin{equation}
  \label{eq:coeff_eff}
 D_{\textrm{eff}}  =  \frac{\gamma a^2}{2 d}
  \left( \frac{\Gamma}{\Gamma+\Lambda}\right)
  \left[ 1 +\frac{2\gamma}{\Lambda} (1-\rho)(1-2\rho) \right].
\end{equation}
The local distribution of cells $\rho(\rbf,t)$ creates implicit
spatial contributions for $D_{\textrm{eff}}$ and thus
(\ref{eq:master_eff}) is classified as a non-linear diffusion
equation. We can check the limit as $\rho$ tends to 0 of the expression
(\ref{eq:coeff_eff}) using the average square displacement $\langle
R^2 \rangle$ for a single particle. In this case, no correlations
appear and the global diffusion coefficient $D$ corresponds
to the spatial average of $D_{\textrm{eff}}$. In practice, the spatial
averaging of (\ref{eq:coeff_eff}) simply replaces the local densities
by the concentration of cells $N'/V$. For a single particle, $N'=N=1$. Indeed,
(\ref{eq:coeff_eff}) is in excellent agreement with $D$ as shown in
Fig.~\ref{fig:single}. 

\begin{figure}
  \includegraphics[width=0.9\columnwidth]{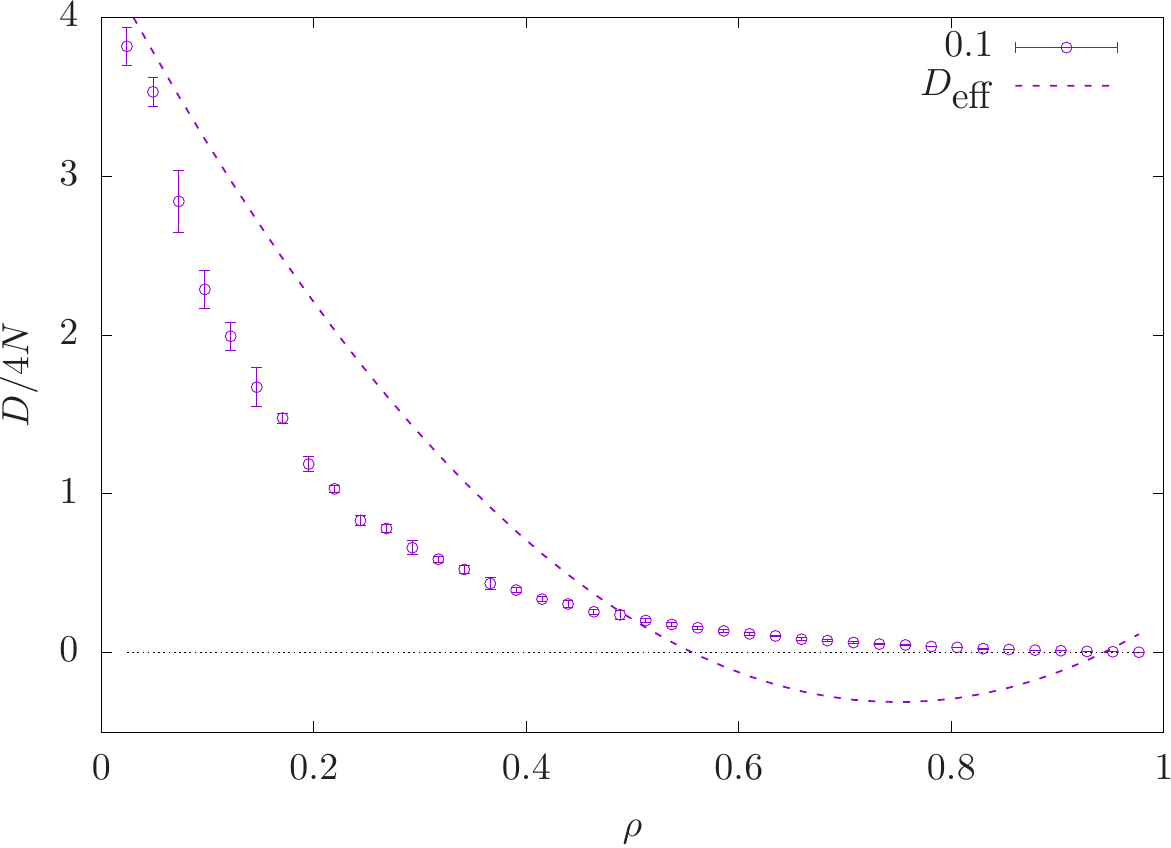}
  \caption{\label{fig:break} Breakage of the mean-field
    approximation. Global diffusion coefficient (circles) extracted
    via the asymptotic least-square fitting  $\langle R^2 \rangle
    \sim 4 D N t$ for $\Lambda/\gamma = 0.1, V=64^2$ and $10^4$ Monte
    Carlo samples. The dashed curve represents $D_{\textrm{eff}}$.}
\end{figure}

Nucleation induced by cellular jams disrupts the flow of cells across
the lattice leading to an overall reduction of migrating cells and
circulating volume for gaseous particles. The accumulation of
particles at nucleation sites creates regions with increased cell
densities in which $D_{\textrm{eff}}$ can take local negative values
(see Fig.~\ref{fig:break} dashed curve).
This means that the local effective cell current
$\mathbf{j}_{\textrm{eff}}(\rbf,t) = -D_{\textrm{eff}} \nablabf\rho$
transport particles to high-density regions, contrary to the effective
pressure caused by cell-cell exclusion. In addition, the average
velocity and dispersion cells in these circumstances are
locally reduced -- they slow down around regions and tend to
accumulate.
This
phenomenon drives the system towards MIPS. From a thermodynamical
perspective, the co-existence of phase creates a concentration
gradient supported by fast moving cells in one region and slow ones in
the other \cite{krauthNatComm2018,lowenPhysRevLett2019}. In a truly 
thermodynamical system, each region (bulk) would be treated as a
particle reservoir with well-defined temperature with negligible
boundary effects (surface). The equilibrium occurs when currents from and into
the gas phase vanishes or, equivalently, when the competing chemical
potentials match each other at phase interfaces.

\begin{figure}
  \includegraphics[width=0.75\columnwidth]{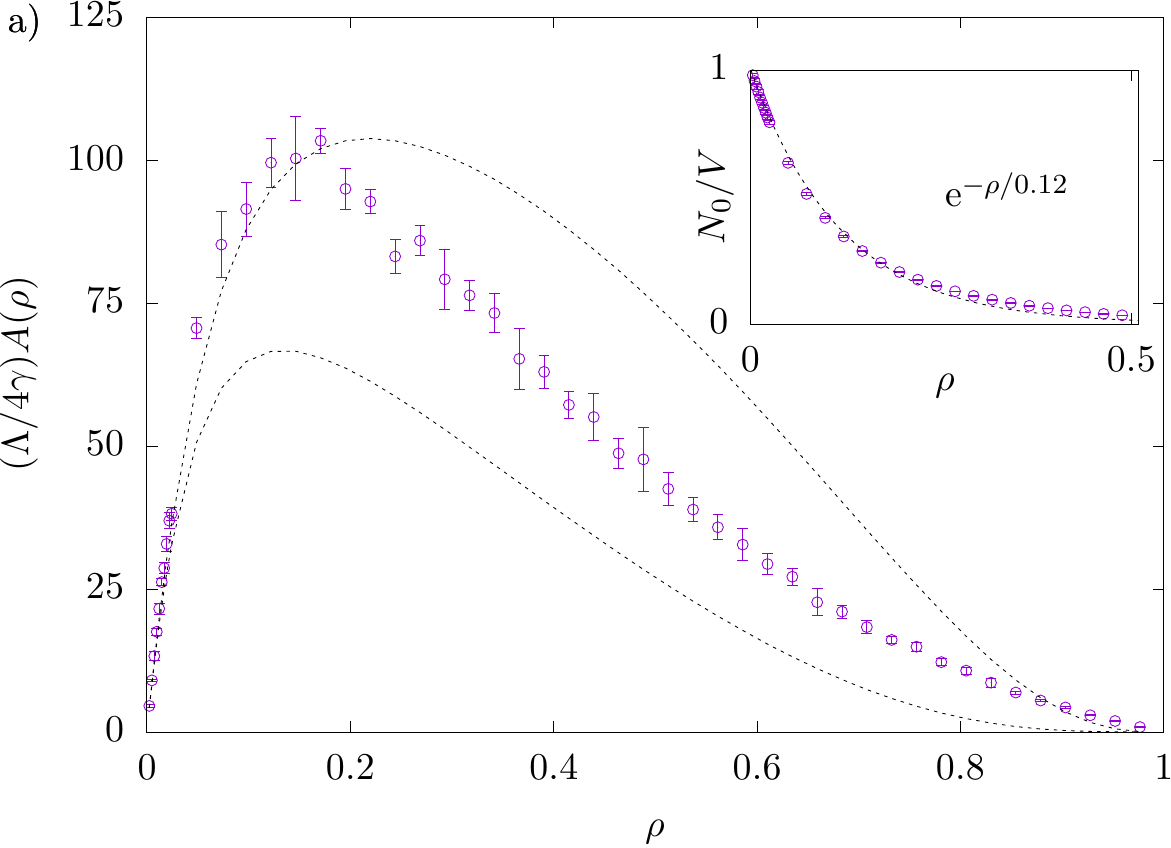}
  \includegraphics[width=0.75\columnwidth]{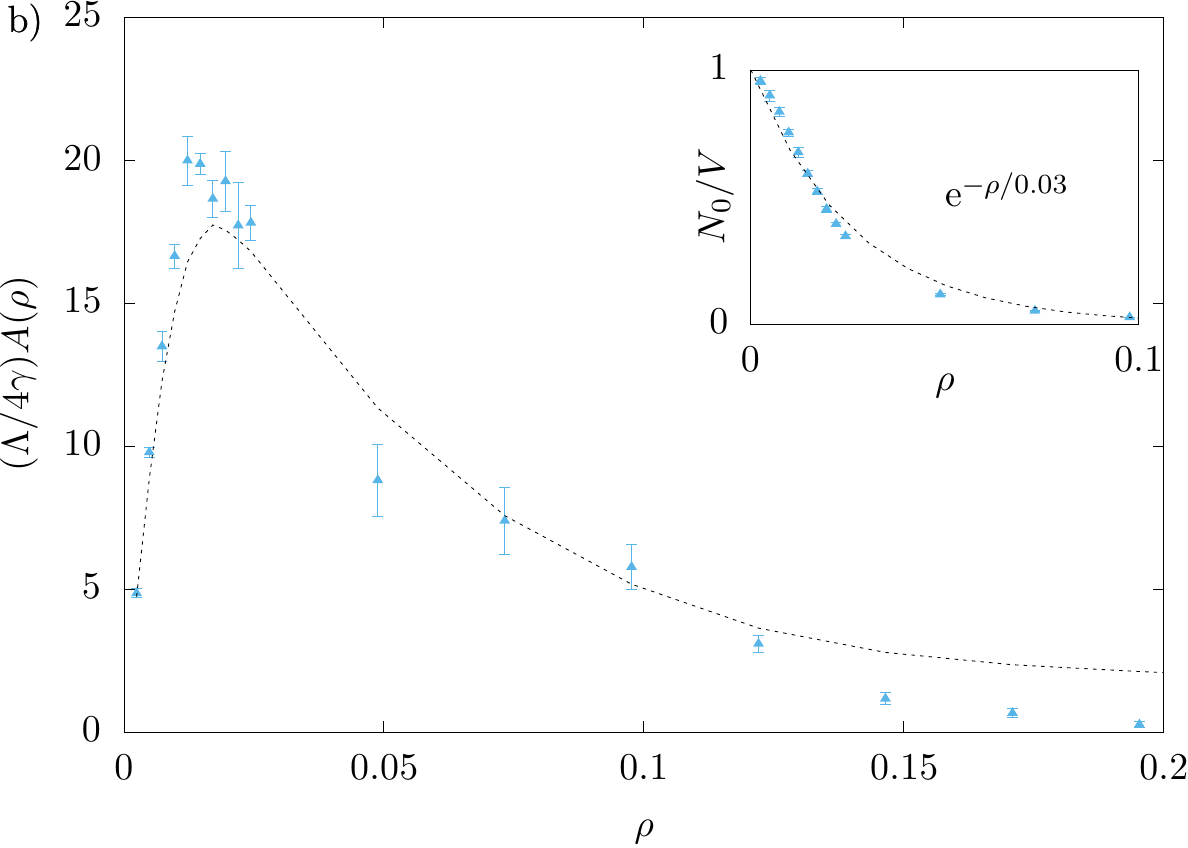}
  \caption{\label{fig:coeffs} Diffusion coefficients for high and low
  depolarization regimes. Scaled angular coefficient $A(\rho)$
  obtained from asymptotic least-square fitting of $\langle R^2 \rangle
  \sim A t $ for various cell densities. The dashed curves represent $(\Lambda/\gamma)
  N' D_{\textrm{eff}}(\rho')$ where $N' = \rho' V$ is the assumed
  number of particles in the gas phase. a) High depolarization regime
  $\Lambda/\gamma=0.1$ and $N'=N_0$ ($N'=N0+N_1/2$) for the bottom
  (top) dashed curve. (inset) density of isolated cells decays
  exponentially. b)  $\Lambda/\gamma=0.001$ and $N'=N_0$.} 
\end{figure}


{
For finite systems, exclusion and spatial correlations introduce
deviations that break the mean-field approximation except for very low
densities (see Fig.~{\ref{fig:break}}). 
In this case, boundary effects cannot be readily dismissed and a
detailed analysis of correlations at gas-liquid interfaces becomes
necessary, which lies well beyond the scope of this study. Instead we
can understand general aspects of the gas phase by considering
different estimates for $N'$ while also neglecting eventual
displacements that might occur within the cluster bulk. 
}

As a first approximation, consider that the gas phase entails
$N'=N_0$ isolated cells, {i.e., cells without any neighbours}. We compare the expected global diffusion $N_0
D_{\textrm{eff}}(N_0/V)$ with the asymptotic angular coefficient $A$
obtained from $\langle R^2 \rangle$, for various cell densities. The
results are shown in Fig.~\ref{fig:coeffs}. For very low densities, we
obtain good agreements for any depolarization regime as expected since
the mean-field approximation holds. As the density increases, collisions
become more prominent and the behavior of the system changes depending
on the depolarization regime. For high depolarization (low
persistence) the curve for $N_0 D_{\textrm{eff}}(N_0/V)$
underestimates $A$ (Fig.~\ref{fig:coeffs}a bottom dashed
line). However, the curve and the data share the same tendencies 
except in the interval of densities near their peaks. In practice this
interval separates two distinct regimes of correlation between the
$N_0 D_{\textrm{eff}}(N_0/V)$ and the data $A$. For lower densities,
the ratio between the curves approaches unity. The ratio in the
remaining interval produces a slow varying function that depends on
$N/V$. This occurs because short-lived small clusters created by
random collisions become more prominent as the density increases and
they are not taken into account by $N_0$. If we include particles with
up to one neighbour $N'=N_0+(1/2)N_1$ then the new curve overestimates
the global diffusion of the system, as indicated by the top dashed
curve in Fig.~\ref{fig:coeffs}a). Here, $N_1$ is the number of
particles with one neighbour and the factor $1/2$ is added 
to account for double counting. The correct value of $N'$ lies in
between both estimates and might include additional corrections from
the particles with 2, 3 and 4 neighbours. Each of category of particle
in respect to the number of cells changes differently according to the
total number of particles in the system. The inset of
Fig.~\ref{fig:coeffs}a) shows how the density of isolated cells
$\rho_0 \approx \mathrm{e}^{-\rho / \rho_c}$ decays with the cell
density with decay constant $\rho_c = 0.12$. In absence of
persistence, the likelihood to find isolated cells should decay as
$\mathrm{e}^{-\rho/2d}$ for very low densities. This suggests that
low persistence can produce observable effects to the aggregation
rate even though the depolarization is not enough to foment stable
nucleation sites.

{
  We stress that our model disregards proliferation, which is what
  grants the malignant aspect of cancer cells. As a consequence, the
  removal of isolated particles indicated in the inset of
  Fig.~\ref{fig:coeffs} is expected at the start of the jamming
  transition. Empirical evidence for transformed MCF-10A mammary
  epithelial cells shows that the same phenomenon is observed in
  real cell systems \cite{leggettPNAS2019}. The removal of dispersed
  individuals and adhesion to clusters results in a reduced
  proliferation rate due to contact inhibition. The coupling of both
  effects, cluster formation with contact inhibition, effectively
  reduces the number of isolated cells.} 

Figure~\ref{fig:coeffs}b) depicts the scaled global diffusion in the low
depolarization regime. In this case, the agreement
between the simulated data and (\ref{eq:coeff_eff}) with $N'=N_0$
(dashed line) is only seen at very low densities $\rho < 0.03$. After
that, the scaled global diffusion quickly approaches zero whereas
$(\Lambda/\gamma)N_0 D_{\textrm{eff}}( N_0/V)$ retains finite
values. Extending $N'=N_0+N_1/2$ do not improve the agreement or
trends. From the inset we can approximate the density
of isolated cells $\rho_0$ by an exponential decay with decay constant
$\rho_c = 0.03$ which means cells tends to strongly aggregate when
compared to random Brownian particles. Unlike the previous case,
and in light of the poor correlation between the scaled global
diffusion coefficient and our estimates, coupled with the rapid
depletion of isolated cells, one can conclude the gas phase ceases to
exists past $\rho_c$. One possible explanation is that the mean-free
path of the isolated cells is much shorted than the average distance
between clusters so that they can be tracked down to cluster
boundaries. The process describes the inter cluster exchange of
isolated cells rather than exchange of cells between two phases in
addition to their natural dynamics.

{ So far, we have not addressed the static properties of the clusters
  nor their implications to the particle dynamics. The static properties
  provide informative clues about the self-organization of particles into
  clusters and they have been extensively explored in the context of
  percolation and diffusion-limited aggregation
  \cite{wittenPhysRevLett1981}. The scale invariance observed for 
  growing clusters also permits the analysis of anomalous scaling
  dimensions for specific metrics, which are often related to the
  types of interactions present in the system. Given the typical
  fractal-like morphology of the clusters (see 
  Fig.~\ref{fig:clusters2}) the scaling relation $n \propto R^{d_f}$
  between the number of individuals in the cluster $n$ and the
  gyration radius $R$ is governed by the fractal dimension $d_f$.

  The fractal dimension of polarized cells at fixed $N/V = 0.2$ and
  $\Lambda/\gamma=0.001$ using the box counting method, $d_f = 1.47\pm
  0.04$. For reference, the fractal dimensions of the 2D 
  self-avoiding walker and diffusion-limited aggregation sit at $4/3$
  and $\sim 1.7$, respectively. In that sense, our model would 
  fall in between two processes with very dissimilar diffusion rules,
  namely, strong repulsion or irreversible adhesion, respectively. On
  grounds of fractal dimensions alone, our model is closer to
  random lattice animals whose fractal dimension is $\sim 1.54$ (Table
  I in \cite{wittenPhysRevLett1981}) for roughly disconnected
  clusters. In the random lattice animal model, clusters with varying
  shapes are placed at random in a lattice to produce configurations
  with fixed connection probability $p$ \cite{petersZPhys1979}. If one
  does not fix $p$, the system converges exponentially towards
  equilibrium characterized by a random distribution of gaps
  \cite{vrhovacPhysRevE2020}. 
  
  One striking feature of polarized particles is their ability to
  aggregate (reversibly). In that regard, the large concentration of
  particles at cluster borders effectively removes isolated ones from
  the cluster surroundings, producing gaps in the spatial distribution.
  In practice, this simulates poorly connected percolation clusters
  ($p\rightarrow 0$). As pointed in \cite{petersZPhys1979}, the random
  lattice model exhibit conceptual similarities with clusters formed
  via Eden process, in which particles are added to a randomly
  selected perimeter site at each time step, starting from a single
  seed particle. Furthermore, the Eden process is also used to justify
  the emergence of non-linear terms in the famous KPZ equation for
  growing interfaces \cite{kardarPhysRevLett1986}. The expansion of
  (\ref{eq:coeff_eff}) for vanishing densities produces
  $\partial_t{\rho} \approx  A_1 \nabla^2 \rho - A_2 | \vec{\nabla}  
  \rho |^2$ which is the averaged KPZ equation with constants $A_1=
  D_0 ( 1 + 2\gamma/\Lambda )$, $D_2= 6 D_0\gamma/\Lambda$, and $D_0 =
  (\gamma a^2/2d )\Gamma/(\Gamma+\Lambda)$. The equation can be
  further simplified using the transformation $\rho = - B \ln W(r,t)$
  with $B = A_1/A_2$ so that one obtains the diffusion equation
  $\partial_t W = A_1 \nabla^2 W$. A noise term $\xi(t)$ can be added
  to describe the average arrival minus departure rate of particles at the
  interface of the clusters and thus recover the complete KPZ
  equation where $d=2$ is the critical dimension with dynamical
  exponent $z=3/2$.

  For larger particle concentration $N/V$, the approximation no longer
  holds and non-linear terms proportional to $\rho$ must be taken into
  account in the scaling analysis. Perhaps, more surprising is that
  the fractal dimension in the numerical simulations does not deviate
  too much from the values expected from the random lattice animals,
  suggesting the corrections are irrelevant for a certain range $(N/V,
  \Lambda/\gamma)$. A far more systematic investigation on the dependence of
  fractal dimensions with global parameters $N/V$ and
  $\Lambda/\gamma$ is beyond the scope of this study. } 

\section{Conclusion}
\label{sec:conclusion}

Active Brownian particles (ABP) systems display a number of unusual
features and tend to 
self-organize into structures that might present relevant biological
properties for living beings. Those features and structures are tied
on how the ABP move and interact with each other and their
surroundings. Among the myriad of types of movement performed by
cells, directional persistence plays a major role in cell
migration. In general, the directional persistence is dictated by 
chemotaxis but different scenarios arise that are not entirely
explained by gradients of external molecules. That includes the
migration of glioma cells that can even travel long distances in the
brain. Although the ultimate goal of any ABP concerns the organization
of biological elements, living beings might contain dozens or even
hundreds of parameters and variables that might affect their
displacement. To keep the analysis as tractable, one needs to reduce
to set of parameters to the bare minimum that encloses the most
significant factors. Here we consider a simple automaton in a square
lattice that accounts for cell exclusion and directional persistence
via the cellular polarization-depolarization process. We derive the
master equation for the distribution of polarized cells and calculate
the effective diffusion coefficient within the mean-field
approximation. Despite the lack of adhesion forces in the model, our
findings show the effective diffusion coefficient can acquire local
negative values producing an effective current that transport cells
from regions with low density of cell to rich ones. The process leads
to self-jamming and favors the creation of nucleation sites, serving
as an example of MIPS between clusters (liquid-like phase) and
isolated cells (gas-like phase).

We also observe two regimes that can be characterized by the
depolarization rate. In the high depolarization regime, self-jamming
do not last long enough to facilitate aggregation. As a result, the
system reaches equilibrium with constant asymptotic value for one-step
square displacement $\langle S^2 \rangle$, and translation invariance
is preserved as demonstrated by the uniform time-averaged distribution
of cells across the lattice. Furthermore, the formula obtained for the
effective diffusion coefficient for all isolated cells captures the
general trends observed for the global diffusion coefficient of the
system, with excellent agreement for low densities due to only small
deviations from the mean-field approximation. Our findings also show
that the directional persistence produces a non-negligible effect on
the decay scale of the isolated cells. 

In the low depolarization regime nucleation centers caused by
self-jamming live long enough to support the organization of cells
into large clusters. These structures can support a large portion of
the cells in the system even at low concentrations. Unlike the
previous regime, our results show an expressive disagreement between
the global diffusion coefficient from the simulated data and our
estimates using isolated cells, except for very low densities.
Isolated cells are thus in-transit particles between the emission and
respective arrival at cluster boundaries, driven entirely by surface
effects and encoded by the escape rate $\varepsilon$.
The phenomenon illustrates the effects of spatial correlations in the
self-organization of ABP, and highlights the weakness in our approach
since fluctuations are discarded in the mean-field approximation.
Thus, a clear explanation of the phenomenology behind surface effects
and spatio-temporal correlations becomes necessary to improve our
understanding on the role played by directional persistence in
organization of active matter.




%

\end{document}